\documentstyle[aps,preprint]{revtex}
\begin{document}
\bigskip
\bigskip
\bigskip
\centerline{\bf CHEMICAL ENRICHMENT AT HIGH REDSHIFTS}
\bigskip
\bigskip
\bigskip
\bigskip
\centerline{SNIGDHA DAS and PUSHPA KHARE}
\centerline{Physics Department, Utkal University}
\centerline{Bhubaneswar, 751004}
\centerline{India}
\newpage

\centerline{\bf ABSTRACT}

We have tried to understand the recent observations related to
metallicity in Ly $\alpha$ forest clouds in the framework of the two
component model suggested by Chiba \& Nath (1997). The model consists
of mini-halos having circular velocities smaller than $\sim$ 55 km
s$^{-1}$, with no star formation and galactic halos with higher
circular velocities $\le$ 250 km s$^{-1}$, having clouds, star
formation and consequent metal enrichment. The number of clouds in the
galactic halos was determined from the observed redshift distribution
of Ly $\alpha$ lines. We find that even if the mini-halos were
chemically enriched by an earlier generation of stars, to have [C/H]
$\simeq$ -2.5, the number of C IV lines with column density
$>$10$^{12}$ cm$^{-2}$, contributed by the mini-halos, at the redshift
of 3, would be only about 10\% of the total number of lines, for a
chemical enrichment rate of (1+z)$^{-3}$ in the galaxies. Not much
information about the degree of enrichment of the mini-halos can
therefore, be obtained from observations of C IV lines with column
density $>$10$^{12}$ cm$^{-2}$. Recently reported absence of heavy
element lines associated with most of the Ly $\alpha$ lines with H I
column density between 10$^{13.5}$ cm$^{-2}$ and 10$^{14}$ cm$^{-2}$
by Lu et al (1998), if correct, gives an upper limit on 
[C/H]=-3.7, not only in the mini-halos, but also
in the outer parts of galactic halos. This is consistent with the
results of numerical simulations, according to which, the chemical
elements associated with the Ly $\alpha$ clouds are formed in situ in
clouds, rather than in an earlier generation of stars.  However, 
the mean value of 7$\times 10^{-3}$ for the column density ratio 
of C IV and H I, determined by Cowie and Songaila (1998) for low 
Lyman alpha optical depths, implies an abundance of [C/H]
=-2.5 in mini-halos as well as most of the region in galactic halos, 
presumably enriched by an earlier generation of stars. The redshift and
column density distribution of C IV has been shown to be in reasonable
agreement with the observations.

\noindent{\it Subject headings}: galaxies: abundances -- galaxies:
halos -- quasars: absorption lines
\section {Introduction}

In the past few years a wealth of information has been obtained, about
the absorbers producing Ly $\alpha$ absorption lines in the QSO
spectra, from the high resolution, high S/N observations obtained with
the Keck telescope as well as observations obtained with the HST. Fernandez-Soto
et al (1997) have used the luminosity function obtained with
photometric redshift measurements of the Hubble deep field galaxies, to
show that all the observed Ly $\alpha$ absorption lines with column
densities $\ge 6\times 10^{13}$ cm$^{-2}$ can be accounted for by
galaxies alone. Important information has come about the presence of
heavy elements in the, till recently considered `pristine', Ly $\alpha$
forest clouds (Tytler et al 1995, Cowie et al 1995). All the Ly $\alpha$ forest clouds with
neutral hydrogen column density,  N$_{\rm H\;I}$, larger than 10$^{15}$
cm$^{-2}$ and 75$\%$ of the clouds with N$_{\rm H\;I}$ $> 3\times  10^{14}$
cm$^{-2}$ at z $\sim$ 3 have been found to show associated C IV
absorption (Songaila \& Cowie 1996). Carbon abundance in these clouds
has been estimated to be between [C/H] $\sim$ -2 and -3. It has been
suggested that a first generation of, Pop III, stars may have polluted
the entire universe to a nearly uniform level of [C/H] $\simeq$ -2.5
(Songaila \& Cowie, 1996, Miralda Escude \& Rees, 1997).  The presence
of heavy elements in lower H I column density clouds has been addressed
recently by Lu et al (1998). They tried to detect C IV in very high S/N
($\sim$ 1860:1) spectra, obtained for a rest frame composite spectrum
of about 300 Ly $\alpha$ lines at redshifts between 2.2 and 3.6. 
In absence of any detection of C IV
lines in the composite spectrum they concluded that the mean
metallicity of the Ly $\alpha$ clouds with 10$^{13.5}$$<$N$_{\rm
H\;I}$$<$10$^{14}$ cm$^{-2}$ is [C/H] $<$ -3.5, at least a factor of 10
lower than that for the higher N$_{\rm H\;I}$ ($>$ 10$^{14}$ cm$^{-2}$)
clouds. Similar conclusion was drawn by Dave et al (1998) who tried and
failed to detect O VI lines  corresponding to weak Ly $\alpha$ lines.
These observations are consistent with results of numerical
hydrodynamical simulation studies of structure formation.  According to
these, most of the metals in Ly $\alpha$ clouds with N$_{\rm H\;I}$ $>$
10$^{14}$ cm$^{-2}$ are produced in situ by Pop II stars in the clouds
themselves or in nearby galaxies, while the first generation of stars
which probably formed around z $\sim$ 14 enriched the universe to a
mean metallicity as small as 10$^{-5}$ solar (Gnedin 1997).

The conclusions of Lu et al (1998) have been contradicted by Cowie 
and Songaila (1998). They have used a new, robust method for analyzing 
the carbon and oxygen absorption lines in the QSO spectra, based on the 
measurement of optical depths. They conclude that the ratio 
N$_{\rm C\; IV}/N_{\rm H\; I}$ roughly remains constant over a wide 
range of over-densities and suggest that a substantial metallicity is 
present even in the regions of the intergalactic medium with only slight 
over-densities. They determine a mean and median value of 7$\times 
10^{-3}$ and $3\times 10^{-3}$ respectively for 
N$_{\rm C\; IV}/N_{\rm H\; I}$, for 
optical depths between 2 and 5 which is close to the range of H I 
column densities considered by Lu et al (1998). The mean value of 
N$_{\rm C\; IV}/N_{\rm H\; I}$ was earlier found to be 3.5$\times 
10^{-4}$ for N$_{\rm H\; I}$ between 10$^{15}$ and 10$^{17}$ cm$^{-2}$
 (Songaila and Cowie 1996).  

In the simulation studies, Ly $\alpha$ absorption with N$_{\rm H\;I}$ $>$
10$^{14}$ cm$^{-2}$ mostly occurs in continuous filaments of gas
surrounding and connecting collapsed objects, while clouds with N$_{\rm
H\;I}$ $<$ 10$^{14}$ cm$^{-2}$ are preferentially found in voids which
are far away from the collapsed objects. As a result, a large variation
is to be found in the heavy element abundances of Ly $\alpha$ clouds.
Clouds with low column densities will have practically no heavy
elements, while the higher column density clouds would have been
enriched by heavy elements from Pop II stars mostly through galaxy
mergers (Gnedin 1997; Gnedin \& Ostriker 1997; Ostriker \& Gnedin
1996). A similar model was recently considered, analytically, by Chiba
\& Nath (1997, hereafter CN97). Their model, based on CDM models of
structure formation, assumes that the Ly $\alpha$ absorption lines are
produced by virialized gas clouds. They have argued that gas confined
to mini-halos (with circular velocities $\le$ 55 km s$^{-1}$) will not
have heavy elements as the star formation in these may be inhibited by
the suppression of radiative cooling due to photoionization. These
mini-halos will contribute more to the lower H I column density Ly
$\alpha$ lines which are not accompanied by lines of heavy elements.
Gas in the galactic halos (55$\le$V$_{c}$$\le$250 km s$^{-1}$), on the
other hand, would probably cool and form clouds and stars and would
contribute more to the higher H I column density Ly $\alpha$ lines
which will be accompanied by lines of heavy elements. It is however,
possible that the mini-halos may have been chemically enriched by an
earlier generation of stars. In this paper we reconsider this two
component model of the Ly $\alpha$ lines and compare its predictions
with the above mentioned observational results.  In section 2 we
briefly describe the model while in section 3 we compare the model
results with observations. Conclusions are presented in section 4.

\section{Description of the basic model}

Our treatment of mini-halos and galactic halos is essentially same as,
and differs only in some details from, that of CN97. We therefore
describe the basic scenario very briefly here, the reader is requested
to refer to CN97 for further details, in particular about the
justification of the values assumed for various quantities entering the
calculation.

We assume the radial distribution of baryonic matter, the density of
which is assumed to be 0.05 times the total matter density in the mini
and galactic halos, as given by
\begin{equation}
\rho (r)\;={\rm V_c^{2}\over{4\;\pi\;\rm
G\;(r^{2}+r_c^{2})}}\;({\rm x_v\over{\rm x_v-\rm arc \;\rm tan\; \rm
x_v}})
\end{equation}
where V$_{c}$ is the circular velocity at the virial radius, r$_{\rm
vir}$, G is the universal gravitational constant, r$_{\rm c}$ is the
core radius, taken to be 10 and 100 kpc for the mini and galactic halos
respectively and $ \rm x_{\rm v} ={\rm r_{\rm vir}\over{\rm r_{\rm
c}}}$. The limits on V$_{\rm c}$ pertaining to the two types of halos
are based on various physical reasons. The lower limit on V$_{c}$
($\sim$ 15 km s$^{-1}$) is roughly the value below which the
perturbations in baryonic mass are suppressed.  The upper limit on
V$_{\rm c}$ ($\sim$ 55 km s$^{-1}$) for mini-halos is roughly the value
below which photoionization suppresses cooling and formation of
clouds.  Upper limit on V$_{\rm c}$ for galactic halos is taken to be
250 km s$^{-1}$. Our treatment of galactic halos differs somewhat from
that of CN97. While CN97 have assumed that the column density along a
line of sight with impact parameter b through a galactic halo is equal
to the column density of hydrogen in a single cloud at a radial
distance equal to the impact parameter from the centre of the galaxy,
we consider the distribution of clouds and calculate the total column
density along a given line of sight as the sum of column densities in
all the clouds that may lie along the line of sight. This is likely to
be more appropriate in view of the covering factors obtained in our
model as will be discussed below. Thus the column density of a
particular ion along a line of sight with impact parameter b is given
by
\begin{equation}
{\rm N_{ion}(b)}\;=2{\rm{\int\limits_{b}^{\infty}{{\rm n_{\rm H}(r)
\rm f_{\rm ion}(r) r dr}\over{\sqrt{r^{2}-b^{2}}}}}} ,\;\;\;\;\;\rm
for\;\rm mini-halos 
\end{equation}
and
\begin{equation}
{\rm N_{ion}(b)}\;=2{\rm{\int\limits_{b}^{\infty}{{\rm N_{\rm H}(r) \rm
f_{\rm ion}(r) \sigma(r)\eta_{cl}(r)rdr}\over{\sqrt{r^{2}-b^{2}}}}}}\;
+\; 2{\rm{\int\limits_{b}^{\infty}{{\rm n_{\rm H}(r) \rm f_{\rm ion}(r)
r dr}\over{\sqrt{r^{2}-b^{2}}}}}},\;\;\;\;\;\rm for \;\rm galactic
\;\rm halos,
\end{equation}
where the first term is the contribution from the galactic clouds while
the second term is the contribution from the intercloud medium along
the line of sight.  Here  n$_{\rm H}$ is the number density of
hydrogen, f$_{\rm ion}$(r) is the ratio of the number density of the
ion to hydrogen number density in a cloud at a radial distance r.
$\sigma$(r) is the area of crosssection of the clouds and $\eta_{\rm
cl}$(r) is the number of clouds per unit volume at a radial distance
r.  We assume all the clouds to have a uniform mass, M$_{\rm cl}$, of
$\sim$ $10^{6}$ M$_\odot$.  The clouds are assumed to be in pressure
equilibrium with the hot intercloud medium. The clouds, being
photoionized, have a temperature $\sim$ $3\times10^{4}$ K. The radius
of a cloud, R$_{\rm cl}$, at a radial distance r from the centre of the
galaxy, in a halo with circular velocity V$_{\rm c}$, is given by
\begin{equation}
\rm R_{cl}\;=({3\rm M_{cl}\over{4\;\pi\rm n_{cl}\rm
m_{p}}})^{1/3}\;=5.2\;\times \;10^{7}\;\rm M_{cl}^{1/3}\;\rm
T_{cl}^{1/3}\;\rm T^{-1/3}\;\rm n^{-1/3}
\end{equation}
T, n and T$_{\rm cl}$, n$_{\rm cl}$ being the temperatures and the
particle densities in the hot medium and the cloud respectively. For
the hot inter cloud medium associated with galactic halos, T for
V$_{c}$ $\ge$ 55 km s$^{-1}$ is given by, T $\simeq\;10^{5}({\rm V_{\rm
c}\over{55 \rm km \rm s^{-1}}})^{2}$ K while for V$_{c}$ $\le$ 55 km
s$^{-1}$ it is taken to be $\sim$ $3\times10^{4}$ K. N$_{\rm H}$(r),
the column density of hydrogen through the cloud at a radial distance
r, is taken to be n$_{\rm cl}$R$_{\rm cl}$.

While CN97 have made the assumption of unit covering factor by the
clouds of the galactic halo projected on to the sky, we assume the
number density of clouds, $\eta_{\rm cl}$(r), to have the same radial
dependence as the total density, so that
\begin{equation}
{\rm \eta_{cl}(r)\;={\eta_{cl}(0)\over{1+r^{2}/r^{2}_{c}}}} 
\end{equation}
The number of lines of a particular ion per unit redshift interval per
unit column density interval per line of sight is given by (assuming
q$_{o}$ = 0.5)
\begin{equation}
{\rm d^{2}N(z,\rm N_{ion})\over{d\rm N_{ion}dz}}\;={{\rm c\over{\rm
H_{o}}}\;(1+\rm z)^{1/2}{\rm{\int\limits_{\rm V_{l}}^{\rm V_{u}}\rm
n(\rm V_{c},z)\;2\;\pi\;\rm b \;{\rm db\over{\rm dN_{ion}}}\; \epsilon
\;\rm dV_{c}}}} ,
\end{equation}
where b is the impact parameter for a halo of circular velocity V$_{\rm
c}$, for which the column density N$_{\rm ion}$ will be obtained for
the particular ion. $\epsilon$ is the fraction of halos that give rise
to absorption lines, for the galactic halos it is the fraction of halos
having sufficient gas to produce absorption lines. For these,
$\epsilon$ is taken to be 0.69, which is the fraction of late type
galaxies (Postman \& Geller 1984). For mini-halos $\epsilon = 1$.
 From the CDM models of structure formation (Mo,
Miralda-Escude \& Rees 1993) the mass and circular velocity of a halo
are related to the comoving radius r$_{o}$ and redshift z by,
\begin{equation}
\rm M\;=\;{4\;\pi\over{3}}\;\rho_{o}\;\rm r_{o}^{3},\;\;\;\;\rm
V_{c}\;=\;1.67\;(1+\rm z)^{1/2}\;H_{o}\;r_{o}
\end{equation}
where $\rho_{o}$ and H$_{o}$, are respectively, the mean density of the
universe and the Hubble constant at the present time.  n(\rm
V$_{c}$,\rm z)dV$_{c}$ is the number of halos per unit volume with
circular velocity between V$_{c}$ and V$_{c}$ + dV$_{c}$ being given by
\begin{equation} \rm n(\rm V_{\rm c},\rm z)\rm dV_{\rm
c}\;=\;{-3(1.67)^{3}\delta_{c}\rm
H_{o}^{3}(1+z)^{5/2}\over{(2\pi)^{3/2}V_{c}^{4}\Delta (r_{o})}}\;{\rm
d\;\rm ln\;\Delta\over{\rm d\;\rm ln\;V_{c}}}\;\times\;\rm
exp({-\delta_{c}^{2}(1+z)^{2}\over{2\;\Delta^{2} (r_{o})}})\;dV_{c}
\end{equation} Here $\delta_{c}$ = 1.68 and the functional form of
$\Delta({\rm r_{o}})$ for the CDM power spectrum of density
perturbation is \begin{equation} \Delta(\rm r_{o})\;=\;16.3\;\rm b_{\rm
g}^{-1}\;(1-0.3909\rm \;r_{o}^{0.1}+0.4814\rm \;r_{o}^{0.2})^{-10}
\end{equation} where b$_{\rm g}$ is the bias parameter taken to be 1.

We have used the code `cloudy 90' for calculation of f$_{\rm ion}$. For
the range of column densities that we are interested in this paper, the
values of f$_{\rm ion}$ for C IV and H I are independent of the
hydrogen column density as well as the particle density for the range
of values expected in the clouds and intercloud medium. ${\rm f_{\rm
ion}\over{Z}}$ is also independent of Z, the chemical abundance of
carbon. We have therefore stored the results of several runs of cloudy
in the form of a table of ${\rm f_{\rm ion}\over{Z}}$ vs. the ionizing
parameter $\Gamma$. The ionization parameter at a radial distance r
from the centre is given by
\begin{equation}
\Gamma(\rm r)\;={{4\;\pi\;10^{-21}J_{21}\over{c\;\alpha_{Q}\;\rm 
h\;n_{\rm H}}}}
\end{equation}
where J$_{-21}$ is the intensity of the UV background radiation in
units of 10$^{-21}$ ergs cm$^{-2}$ s$^{-1}$ Hz$^{-1}$ str$^{-1}$. c and
h are velocity of light and Planck's constant and $\alpha_{Q}$ is the
slope of the spectrum of the UV background radiation (assumed to be a
power law). n$_{\rm H}$ is replaced by n$_{\rm cl}$, the density of the
particles in the clouds at a radial distance r for clouds in galactic halos. The
table is searched and the value of f$_{\rm ion}$ obtained by
interpolation, at every point during the evaluation of the integrals in
equation (2) and (3).  J$_{\nu}$ is assumed to be independent of z for
z$>$2.5 as suggested by the proximity effect analysis of Ly $\alpha$
lines, while for z$<$2.5, J$_{\nu}$ varies with z as
J$_{\nu}({1+z\over{3.5}})^\alpha$.  The index $\alpha$ gives the rate
of decrease of J$_{\nu}$ for z$<$2.5 and we have considered two cases
(i) $\alpha$ = 0 (ii) $\alpha$ = 2.0.

The number of clouds per unit volume at the centre of the galaxy,
$\eta_{\rm cl}$(0), is an unknown factor. We take $\eta_{\rm cl}$(0) =
f ${3\over{4\;\pi\;\rm r^{3}_{cl}(0)}}$, where f$\le$1. The value of f
is obtained by making the predicted redshift distribution for Ly
$\alpha$ lines match with the observed values. These distributions are
plotted in Figure 1 for two values of $\alpha$, for $\alpha_{Q}$ =
1.5.  The straight line fit to the observed data from Keck and HST is
from Kim et al (1998). A good match between theoretical and observed
distributions is obtained for f=0.0035. The Figure also shows the
distribution for f = 0.001 and f = 0.01 for comparison. f = 0.0035
gives the total number of clouds in the galactic halos with V$_{c}$ =
55 \& 250 km s$^{-1}$ to be about 380 and 161544 respectively, thus
giving total halo mass $\simeq 3.8\times 10^{8}$ and $1.6\times
10^{11}$ M$_\odot$ which is close to 10 $\%$ of the total masses of
these halos. The total covering factor presented by the clouds on the
projected plane of the galaxy being 0.16 and 1.15 for V$_{c}$ = 55 \&
250 km s$^{-1}$ respectively. In the following analysis we take the
value of f to be 0.0035. Note that CN97 have assumed a unit covering
factor by clouds in the galactic halos, while we get lower values for
low V$_c$ clouds and higher values for high V$_c$ clouds.  As a result,
their assumption of the H I column density at an impact parameter b
being equal to the column density in a cloud at a radial distance b is
not valid in our case.
\section{ Comparison with observations }
\subsection{ Column density distribution of neutral hydrogen}
The column density distribution function, f(N$_{\rm H\;I}$), is defined
as the number of absorbing systems per unit column density per unit
redshift path which is defined by X(z) = ${2\over{3}}[(1+z)^{3/2} - 1
]$ for q$_{o}$ = 0.5. Recently Kim et al (1998) have analyzed the
properties of low column density (between $10^{12.8}$ and $10^{16}$
cm$^{-2}$) Ly $\alpha$ lines towards 5 QSOs at different redshifts,
between 2.1 and 3.5, using the Keck data. They find that the density
distribution function fits a power law with a slope between -1.35 and
-1.55. The slope changes slowly with redshift, the distribution
becoming steeper with increasing redshift. For N$_{\rm H\;I}$ $>$
10$^{14}$ cm$^{-2}$, they find a departure from the power law, the
observed number of lines being smaller than that given by the power
law. A comparison of the observations with the predictions of our model
is presented in Figure 2. The slope of the predicted distribution is
$\sim$ -1.5 and it is almost independent of redshift. The distribution
function also shows an increase in the slope at higher N$_{\rm H\;I}$,
which is, however, less than the observed departure from the power
law.
\subsection{ Redshift distribution of C IV lines}
In trying to determine the redshift distribution of heavy element
lines, one has to consider the change in abundance of heavy elements
with redshift. Direct observational evidence for increase in chemical
abundance with redshift has been obtained, for damped Ly $\alpha$
systems, by Pettini et al (1995, 1997) and Lu et al (1996). Lu et al
(1996) have shown that the mean metallicity of these systems increases
with decreasing redshift. They find that most systems with z $>$ 3 have
[Fe/H] $<$ -2 while at z $<$ 3 a large fraction of the systems have
[Fe/H] $\simeq$ -1.

It thus seems that the abundance in galactic disks ( of which the
damped Ly $\alpha$ are believed to be progenitors (Prochaska \& Wolfe
1997, however, see Pettini et al 1997)) has been increasing with time.
The abundance of carbon in Ly $\alpha$ forest clouds with N$_{\rm
H\;I}$ $>$ 10$^{14}$ cm$^{-2}$ at z $\sim$ 3 has been found to be
between 10$^{-2}$ and 10$^{-3}$ of solar abundance. Though the
abundance of silicon w.r.t. carbon in Ly $\alpha$ clouds appears to
have changed with redshift (Songaila \& Cowie 1996), there is, as yet,
no definite evidence for a change in abundance of carbon with redshift
in the forest clouds. However, it seems possible that the abundance in
the galactic halos has been increasing continuously with time due to in
situ star formation (Khare \& Rana 1993; CN97). We, therefore, assume
the abundance of carbon to depend on the redshift as Z(z) =
Z(0)(1+z)$^{-\delta}$ in the galactic halos. For mini-halos we have
assumed a constant abundance of [C/H]$\simeq$-2.5, assumed to be
produced by the Pop III stars. In Figure 3 we have plotted the results
of calculation for the redshift distribution of C IV lines for both
galactic halos and mini-halos together for N$_{\rm C\;IV}$ $>$ 10$^{13}$
cm$^{-2}$.

In order to obtain observed value of dN/dz, we have collected data from
the literature and performed a maximum likelihood analysis. The results
of this analysis along with the references for the data used are given
in Table 1. The data include components of damped Ly $\alpha$ systems
also. We have included these in the data as most of the components are
likely to arise in the galactic halos. Inclusion of the few pure DLA
components will change the values of ${\rm dN \over{\rm dz}}$ only by a
very small amount. We have also considered a poissonian sample
constructed for the data, obtained by counting lines within 200 km
s$^{-1}$ of each other as one line. This is to count the lines formed
by various clouds in a single galaxy as one line. The results for this
data set are given in the last two lines in Table 1.  The value of the
evolutionary index $\gamma\;({\rm dN \over{\rm dz}}\;\alpha\;(1+\rm
z)^{\gamma})$ is positive if lines with smaller column densities are
included. The index is negative for higher column density lines,
indicating an increase in the number of such lines with decreasing
redshift, similar to that found with low resolution observations
(Steidel 1990). The values of dN/dz for N$_{\rm C\;IV}$ $>$ 10$^{13}$
cm$^{-2}$ are plotted in Figure 3. Songaila (1998) has reported
observations of C IV towards 13 QSOs. The value of dN/dz calculated for
her data with an average redshift of 2.87 is also plotted in the
figure.

As seen from the Figure, high values of $\delta$, $\simeq$ 4, are
indicated by the observations which are probably consistent with the
observations of damped Ly $\alpha$ systems. The distribution shows
negative values of the index of redshift distribution, $\gamma$, for
redshifts $>$ 1. The values of $\gamma$ for N$_{\rm C\;IV}$ $>$
10$^{13}$ cm$^{-2}$ in Table 1 are positive. However, $\gamma$ is
negative for higher column density cutoffs, the values though, are
higher than the values obtained here for $\delta$ = 4.

One important result that emerges from this is that the contribution of
mini-halos, assuming the abundance in these halos to be [C/H]=-2.5,
produced by an earlier generation of stars, to the C IV lines with
column density $>$10$^{13}$ cm$^{-2}$ is completely negligible, being
$<5\%$ at z=3. Even for C IV lines with column density $>$ 10$^{12}$
cm$^{-2}$ mini-halos contribute only about 10 $\%$ to the total number
of C IV lines at z=3. In the galactic halos, the contribution to the
C IV column densities comes mostly from the clouds, the hot intercloud
medium contributing insignificantly to the total C IV column density
due to its lower particle density. Thus even if mini-halos had a
chemical enrichment of [C/H]=-2.5 due to Pop III stars, they would not
contribute significantly to the observed number of carbon lines with
column density $>$ 10$^{12}$ cm$^{-2}$.
\subsection{ Column density distribution of C IV lines}
The column density distribution function of C IV clouds at 2.52 $<$ z
$<$ 3.78 has recently been obtained from high resolution Keck
observations by Songaila (1998). The distribution is roughly a power
law with a slope of -1.5 for column densities larger than
$6\times10^{12}$ for which her sample is complete. We have plotted
model results for the column density distribution at z=3 for C IV, for
$\delta$ = 3 and 4 in Figure 4. The model results bracket the observed
values for column densities $>10^{13}\;\rm cm^{-2}$. However, at lower
column densities the model predicts many more lines than the observed
number. The observed data, however, may be incomplete below
$6\times10^{12} \;\rm cm^{-2}$ as noted by Songaila (1997) and it is
possible that the number of small column density lines may actually be
considerably larger. Note that large values of $\delta$ $\simeq$ 4,
indicated by the dN/dz data, do not produce sufficient number of lines
with column densities $>$ 10$^{13.5}$ cm$^{-2}$. If the number of lines
with column densities $<\;6\times10^{12}$ cm$^{-2}$ is indeed small,
some basic assumptions have to be modified in our model. Increase in
rate of abundance evolution does not serve the purpose as it reduces
the number of higher column density lines more than that of the lower
column density lines. This is discussed further in the next section.
\subsection{ Metal lines associated with low H I column density lines }
As mentioned above, Lu et al (1998) have tried to detect C IV lines
associated with H I lines having column density between 10$^{13.5}$
cm$^{-2}$ and 10$^{14}$ cm$^{-2}$, (referred to below as 'associated C
IV lines'), between redshifts 2.2 and 3.6, by obtaining a composite
rest frame spectra of about 300 lines which did not have accompanying C
IV lines. Note that in a spectra covering a total redshift path of
$\Delta$z = 4.15 (leaving out the region of relative velocity $\le$
5000 km s$^{-1}$ from the QSOs which may be contaminated by the lines
associated with the QSOs) they detected 7 Ly $\alpha$ lines with column
density between 10$^{13.5}$ cm$^{-2}$ and 10$^{14}$ cm$^{-2}$, which
were accompanied by C IV lines. They failed to detect any C IV line in
the composite spectra and obtained an upper limit of 10$^{10.5}$
cm$^{-2}$ on the average column density of the 'associated C IV
lines'.  As noted before, this conclusion is likely to be erroneous
 (Cowie and Songaila, 1998). Here, however, we consider the consequences 
 of this conclusion. The results quoted below can be easily scaled 
for the correct value of the upper limit on the C IV column density. 
 We have also discussed the implications of the results of Cowie and 
Songaila (1998) below. 

We have calculated the range of impact parameters as a
function of circular velocity, which produce Ly $\alpha$ lines with H I
column densities between 10$^{13.5}$ cm$^{-2}$ and 10$^{14}$
cm$^{-2}$.  These are plotted in Figure 5 for the redshift of 3. Note
that the column density range is not obtained for galactic halos with
$\rm{V_{c}}$ $<$ 90 km s$^{-1}$ (even though it is obtained in the
mini-halos), in spite of the fact that we have taken the contribution
from the hot (intercloud) medium into account. This is due to the
higher value of core radius used for the galactic halos. The number of
such Ly $\alpha$ lines per line of sight per unit redshift interval at
z $\sim$ 2, 3 and 4 bracketing the redshifts of the Ly $\alpha$ lines
considered by Lu et al (1998) is given in Table 2. The number of
'associated C IV lines' per unit redshift interval, per line of sight,
is same as the number of Ly $\alpha$ lines. The range of column
densities of the 'associated C IV lines' is plotted in Figure 6 as a
function of the circular velocity for the redshift of 3. It can be seen
that all the 'associated C IV lines', including those due to the
mini-halos, have a column density $\ge$ 10$^{11}$ cm$^{-2}$ and hence
would have been clearly observed by Lu et al (1998). The detection of
only 7 lines by them in a path length of $\Delta z = $4.15 could then be used to
obtain an upper limit on the chemical enrichment due to the Pop III
stars. We find that the abundance  has to be less than -3.7 w.r.t. the
solar value for the column densities of 'associated C IV lines'
produced by the mini-halos to be below 10$^{10.5}$ cm$^{-2}$. The column
density range for mini-halos for this value of abundance at z=3 is shown
in Figure 6 (band 5). This conclusion is similar to that of Lu et al (1998).

The column densities of the 'associated C IV lines' produced by
galactic halos will reduce if the chemical enrichment due to star
formation is confined only to the inner parts of the galactic halos,
the outer parts having an abundance (of -3.7) same as the mini-halos.  This is
expected if further enrichment in these halos is produced due to in
situ star formation, which will occur more efficiently towards the
centres of the galaxies due to the higher densities occurring there. We
therefore, considered two possibilities (i) an abundance gradient
Z(r,z)=Z(0,z)e$^{-2\;\rm {r/V_{c}}}$ for the heavy elements produced in
situ in the galaxies and (ii) an upper limit on the radial distance up
to which heavy elements enrichment has occurred in the galaxies, given
by r$_{\rm max}$ ($\rm{V_{c}}$,z)= ${4\rm V_{c}\over{(1+z)}}$. Here r
and V$_{\rm c}$ are in units of kpc and km s$^{-1}$ respectively. The
column densities of the 'associated C IV lines' for these two
assumptions are also shown in Figure 6 for z=3.  The column densities
of C IV lines are below the detection limit of Lu et al (1998) for most
of the range of circular velocities for both the possibilities. The
number of all these 'associated C IV lines' per unit redshift interval
per line of sight for the assumption (ii) above, is $\ge$ 6.2 so that
for the sample of Lu et al (1998) about 26 lines are expected, some of
which may have column densities below the observable limit. Note that
they have observed 7 lines. The dN/dz for the assumption of abundance
gradient, on the other hand, is same as that in Table 2. However, as
seen from the Figure, most of these would be below the sensitivity of
detection, consistent with observations of Lu et al (1998).

We have calculated the column density distribution of C IV lines for
these two possibilities which is shown in Figure 4 for $\delta $ = 3.0
at z=3. It can be seen that the assumption of abundance gradient
reduces the number of high column density lines much below the observed
number. The assumption of maximum radial distance for in situ chemical
enrichment, however, gives a distribution closer to the observed
distribution. We have also calculated the redshift distributions for
the two possibilities. These are plotted in Figure 3 for $\delta$ = 3
and N$_{\rm C \rm IV}\;>\;10^{13}\;\rm cm^{-2}$. The distribution for
the assumption (ii) above is closer to the observed values. $\gamma$ is
again negative for higher redshifts, its value ($\simeq$ -1.6, for
z$>$2) is somewhat smaller than the values obtained for higher column
density cutoffs in Table 1 and than the value of -1.2 obtained for low
resolution observations by Steidel (1990).

We now discuss the implications of the results of Cowie and Songaila 
(1998) for our results. Taking their mean value of 7$\times 10 ^{-3}$ 
for N$_{\rm C\; IV}$/N$_{\rm H\; I}$, the range of column density of the '
associated C IV lines' is 2.2$\times 10^{11} - 7\times 10^{11}$ cm$^{-2}$.
The lines produced by mini-halos for the assumed carbon abundance of 
 [C/H]= -2.5 (band 5 in Figure 6) are roughly consistent with this range. 
The C IV lines produced by the galactic halos for $\delta = 3$ (band 1 
in Figure 6), have higher column densities, indicating lower or absence of 
chemical enrichment in the outer parts of the galactic halos due to 
in situ star formation (assumptions (i) and (ii) above). We have plotted 
in Figure 6 (band 6) the range of C IV column density for the galactic 
halos assuming a uniform abundance of [C/H]=-2.5. These are consistent 
with the expected range of 2.2$\times 10^{11} - 7\times 10^{11}$ cm$^{-2}$.
 Thus an abundance of -2.5 in the mini-halos as well as in the outer parts 
of the galactic halos due 
to an earlier generation of stars seems to be consistent with the results 
of Cowie and Songaila (1998). The chemical enrichment due to in situ star 
formation has to be restricted to the central regions of the galactic halos
at the redshift of about 3. We have calculated C IV column density distribution
at z=3, $\delta$=3, for the assumption (ii) above, taking the abundance in the
mini-halos and outer parts of galactic halos to be -2.5. This is shown in 
Figure 4. The distribution is close to that for $\delta$=4, for low column
densities $\le 10^{13}$ cm$^{-2}$ as the abundance for $\delta$=4 for z=3 is close to
-2.5. The number of these lines is thus considerably higher than the observed
values. For this possibility the number of C IV lines with column density 
$\ge 10^{13}$ cm$^{-2}$ at z = 2 is 15.8, somewhat larger than the observed 
values.  
\section{Discussion and Conclusions}
We have tried to understand the recent observations of Ly $\alpha$
forest lines and accompanying C IV lines, in the framework of
hierarchical structure formation model. The observed redshift
distribution of the Ly $\alpha$ lines has been used to fix the number
of clouds in galactic halos. The predicted column density distribution
of Ly $\alpha$ clouds is found to be similar to the observed
distribution. We find that at redshifts $\le$ 2, the number of Ly
$\alpha$ lines with column density $>\;6\;\times\;10^{13}\; \rm
cm^{-2}$ contributed by the mini-halos is $\simeq$ 25 $\%$ of the total
lines. Recently Fernandez-Soto et al (1997) have determined the expected density
of crossings for an arbitrary line of sight ( for z $\le$ 2 ) by the
halos of galaxies, the luminosity function for which was obtained by
them from the observations of the Hubble deep field galaxies. They
concluded that this number is consistent with the observed number of Ly
$\alpha$ lines. They, therefore, suggest that all the observed Ly
$\alpha$ lines with N$_{\rm H\;I}$ $>\;6\;\times\;10^{13}\;\rm cm^{-2}$
are produced by lines of sight crossing galactic halos alone and that
no other population ( e.g. the intergalactic clouds ) is needed to
explain the occurrence of these lines. Our values of fraction of Ly
$\alpha$ lines with N$_{\rm H\;I}$ $>\;6\;\times\;10^{13}\; \rm
cm^{-2}$ contributed by mini-halos are, however, not consistent with
these findings.

In this scenario, the Ly $\alpha$ forest lines should show clustering
over velocity scales of a few hundred km s$^{-1}$, over which the
galaxies are known to cluster strongly. Clustering on smaller scales
will also be present due to the multiple clouds crossing the line of
sight in a single halo. Clustering among the forest lines has been
observed (Srianand and Khare, 1994). Recently Fernandez-Soto et al
(1996) studied the clustering by studying the two point correlation
function of C IV lines associated with Ly $\alpha$ lines. They
concluded that Ly $\alpha$ lines with H I column density $> 3 \times
10^{14} \; \rm cm^{-2}$ are strongly correlated in redshift on velocity
scales $\le$ 250 km s$^{-1}$. The clustering seems to persist to lower
column densities, though it seems to be weaker compared to that at 
higher column densities.
 
Chen et al (1997) have tried to investigate the relation between
properties of Ly $\alpha$ absorptions systems and the associated
galaxies. They find that out of a sample of 33 galaxies, 7 do not
produce Ly $\alpha$ absorption. The covering factor of absorbing clouds
in galaxies producing Ly $\alpha$ lines with equivalent width $>$
0.3$\AA$ (N$_{\rm H\;I}$ $\ge\;6\;\times\;10^{13}\; \rm cm^{-2}$) was
earlier found by Lanzetta et al (1995) to be 1 for impact parameters
$<$ 160 kpc while it was significantly smaller for higher impact
parameters. The covering factor for such clouds in galactic halos in
our model for impact parameter $<$ 160 kpc, is 0.03 for z=1 and 0.02
for z=2. However, we have assumed the galactic halos to have circular
velocities from 55 to 250 km s$^{-1}$. It is very likely that the
galaxies observed by Lanzetta et al (1995), on which their estimates of
covering factor are based, may only correspond to circular velocity
towards the higher end of the range considered by us. We have
calculated the covering factor by restricting the circular velocities
to above 100, 150 and 200 km s$^{-1}$, the values are 0.14, 0.34 and
0.43 respectively. These are still considerably smaller than the values
found by Lanzetta et al (1995).

We have shown that the mini-halos would not contribute significantly to
the number of C IV lines with column density $> 10^{12}$ cm$^{-2}$ even
if the heavy element abundance in these halos was [C/H]$\simeq$-2.5.
Thus no definite conclusions about the level of enrichment by Pop III
stars can be drawn from observations of such lines. The reported presence of
very few lines of C IV associated with the Ly $\alpha$ lines with
column density between 10$^{13.5}$ cm$^{-2}$ and 10$^{14}$ cm$^{-2}$, if 
correct, indicates an upper limit of [C/H]$\le$-3.7, not only for the mini-halos,
but also for the outer parts of the galactic halos. We have shown that
heavy element enrichment beyond [C/H]=-3.7, in the galactic halos,
should be confined only to the inner regions of the galaxies in order
to be consistent with the results of Lu et al (1998).
This is shown to be
consistent with the observed distribution of C IV lines. The values of
r$_{\rm max}$, suggested here are larger than the impact parameters
obtained for heavy element line producing galaxies (Bergeron \& Boisse,
1988; Steidel, 1995). The values are also consistent with the expected
distances traveled by the material ejected by supernovae in about
10$^9$ years assuming the velocity of the ejecta to be a few hundred km
s$^{-1}$.  These results are consistent with the suggestions of
numerical simulations studies that the heavy elements observed in the
Ly $\alpha$ absorbers are indeed produced in the absorbing clouds
themselves. The values obtained by  Cowie and Songaila (1998) for the
column density ratio of C IV and H I, however, indicate an 
enrichment of mini-halos as well as galactic halos to [C/H] =-2.5 by 
an earlier generation of stars.

\centerline{\bf Acknowledgement}

This work was partially supported by a grant (No. SP/S2/013/93) by the
Department of Science and Technology, Government of India.

\newpage
\noindent{\bf Figure Captions}
\noindent {\bf Fig 1:} Redshift distribution of Ly $\alpha$ lines with column
density greater than 10$^{13.77}$ cm$^{-2}$. Upper and lower solid
lines are for f = 0.0035, for $\alpha \;=\;2$ and 0 for z $<$ 2.5
respectively. Long dashed line is the best fit line for observed values
taken from Kim et al (1997). Upper and lower short dashed lines are
for $\alpha$ = 0, for f = 0.01 and 0.001 respectively.\\
\noindent  {\bf Fig 2:} Column density distribution of Ly $\alpha$ lines for
$\alpha$ = 2.0. Solid and dashed lines are at z=2.31 and z=3.35
respectively. Stars and solid triangles are observed values  at z=2.31
z=3.35 respectively. \\
\noindent {\bf Fig 3:} Redshift distribution for C IV lines with N$_{\rm C
\rm IV}\;>\;10^{13}\;\rm cm^{-2}$. Small-dashed, solid and long-dashed
lines are for $\delta$ = 0, 3 and 4 respectively. Dash-dotted line is
the distribution obtained with the assumption of an upper limit on
radial distances for which chemical enrichment has occurred while long
and short dashed line is for the case of an abundance gradient being
present in the galactic halos ($\delta$ = 3, for both cases), assuming 
an abundance of -3.7 due to an earlier generation of stars. Triangle
shows the observed value for the data set collected from the literature
(Table 1), circle shows the observed value for the poissonian sample
(obtained by counting lines within 200 km s$^{-1}$ of each other as one
line), while square represents the value for the data from Songaila
(1998).\\
\noindent  {\bf Fig 4:} Column density distribution for C IV lines at z =
3.0. Solid and long-dashed lines are for $\delta$ = 3 and 4
respectively.  Small-dashed and dash-dotted lines are for the
assumptions of upper limit on radial distance for galactic chemical enrichment
and abundance gradient respectively, assuming an abundance of -3.7 in
the mini-halos and outer parts of galactic halos.
Dotted line is for the assumption of upper limit on the radial distance
for galactic chemical enrichment for an abundance of -2.5 in the mini-
halos and in the outer parts of galactic halos. $\delta$ = 3 for all
the three cases. 
Triangles and squares are the observed values from Songaila (1998) for
z$<$ 3 and z $\ge$ 3 respectively.\\
\noindent  {\bf Fig 5:} Range of values of impact parameter for which the
column density of H I is between 10$^{13.5}$ cm$^{-2}$ and 10$^{14}$
cm$^{-2}$ as a function of circular velocity.\\
\noindent  {\bf Fig 6:} Range of column density of C IV for lines of sight
giving rise to Ly $\alpha$ lines with column densities between
10$^{13.5}$ cm$^{-2}$ and 10$^{14}$ cm$^{-2}$ as a function of circular
velocity. Bands labeled 1, 2 and 3 are results for assumptions of
uniform chemical abundance in a given galactic halo ($\delta = 3$), 
radial abundance
gradient in the galactic halos and an upper limit on radial distances
inside galactic halos for heavy element enrichment respectively.  Bands
4 and 5 are for mini-halos with [C/H]=-2.5 and [C/H]=-3.7
respectively,for $\delta$ =3 at redshift of 3. Band 6 is for galactic halos
assuming a uniform abundance of -2.5, at z=3.\\

\newpage
\begin{table}
\caption {Results of maximum likelihood analysis for C IV lines}
\bigskip
\begin{tabular}{cccccc}
\hline
\multicolumn{1}{c}{\rm N$_{\rm C\;\rm IV}^{\rm min}$}&
\multicolumn{1}{c}{No of lines}&
\multicolumn{1}{c}{$\gamma$}&
\multicolumn{1}{c}{$\rm dN\over{\rm dz}$}&
\multicolumn{1}{c}{z$_{\rm av}$}\\
\hline
12.0  &  186.0  &  0.52$\pm$0.42   &  15.5   &  2.47 \\
12.5  &  175.0  &  0.41$\pm$0.43    &  14.6  &  2.46 \\
13.0  &  143.0  &  0.24$\pm$0.48  &  11.9   &  2.45 \\
13.5  &  87.0   &  -1.02$\pm$0.59   &  7.2   &  2.32 \\
13.77 &  58.0   &  -1.22$\pm$0.47   &  4.8   &  2.29 \\
13.0  &  80.0   &  0.78$\pm$0.65    &  6.7   &  2.51 \\
13.5  &  56.0   &  -0.25$\pm$0.76   &  4.7   &  2.39 \\
\hline
\end{tabular}

Ref: Cowie et al 1995, Khare et al 1997, Tripp et al 1996,
Savaglio et al 1996, Petitjean \& Bergeron 1994, Petitjean,
Rauch, Carswell 1994, Tripp, Lu, Savage 1996, Giallongo
et al 1993, Cristiani et al 1995, Williger et al 1994,
Kulkarni et al 1996, Carswell et al 1984, Carswell et al 
1991, Wampler et al 1993, Wampler et al 1991, Petitjean
\& Bergeron 1994, Pettini et al 1983\\
\end{table}
\begin{table}
\caption {Number of Ly $\alpha$ lines with column density between
10$^{13.5}$ and 10$^{14}$ cm$^{-2}$}
\bigskip
\begin{tabular}{cccc}
\hline
\multicolumn{1}{c}{z}&\multicolumn{1}{c}{mini-halos}&
\multicolumn{1}{c}{galactic halos}&\multicolumn{1}{c}{Total}\\
\hline
2 & 14.4 & 65.4 & 79.9\\
3 & 40.5 & 91.9 & 132.4\\
4 & 73.3 & 133.0 & 206.3\\
\hline
\end{tabular}
\end{table}

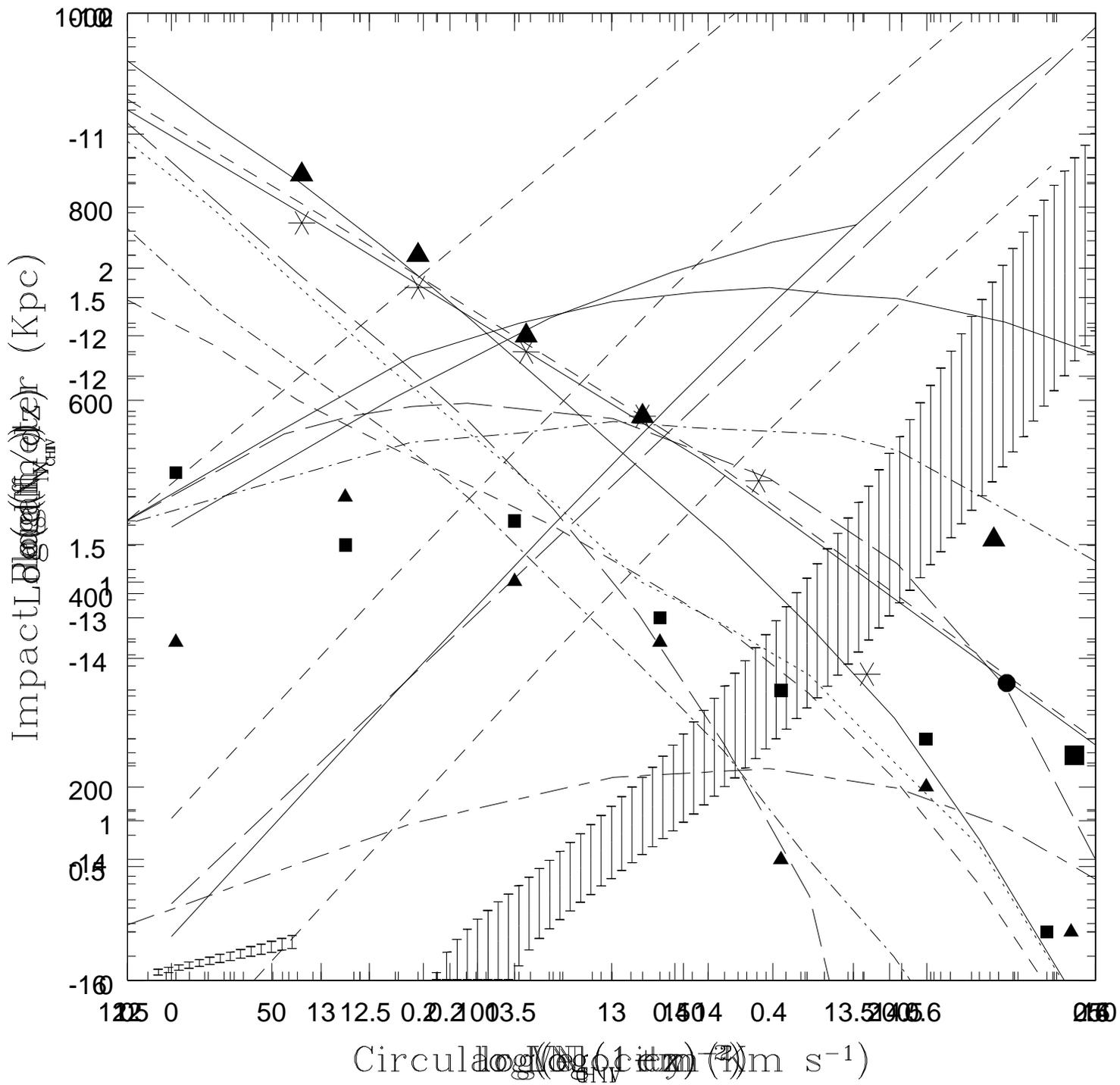
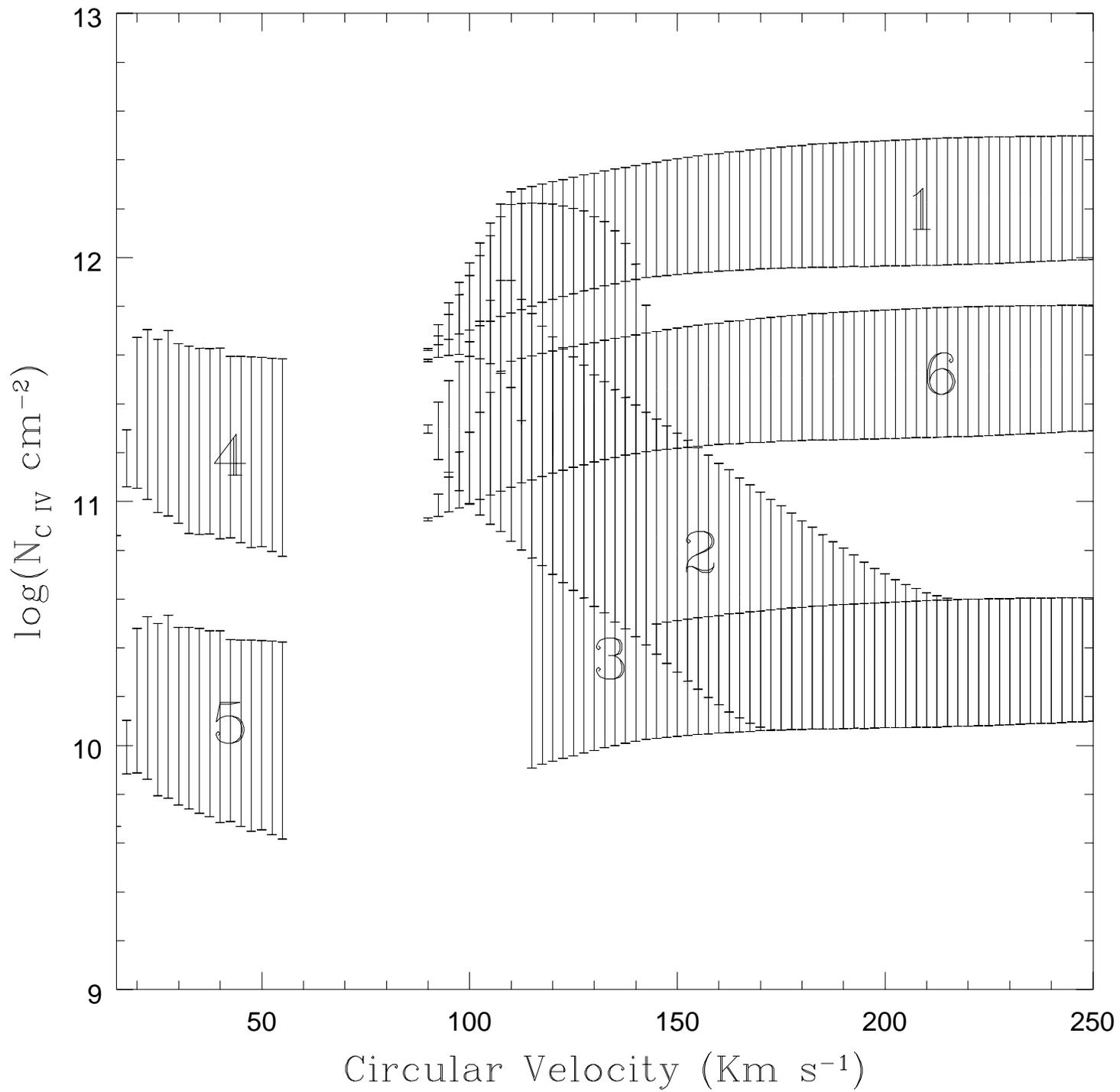
\end{document}